\documentclass[debug,overfull]{epl}

\usepackage{graphics}
\usepackage{epsfig}
\usepackage{amsmath}

\title{Thermal activation of rupture and slow crack growth in a model of homogenous brittle materials}
\shorttitle{Thermal activation of rupture and slow crack growth}
\author{S. Santucci\inst{1} \and L. Vanel\inst{1} \and A. Guarino\inst{2} \and R. Scorretti\inst{3} \and S. Ciliberto\inst{1}.}
\shortauthor{S. Santucci \etal}
\institute{
  \inst{1} Laboratoire de physique, CNRS UMR 5672,
  Ecole Normale Sup\'erieure de Lyon,
  46 all\'ee d'Italie,
  69364 Lyon Cedex 07, France \\
  \inst{2} Universit\'e de Polyn\'esie Fran\c{c}aise, BP 6570 FAA'A, Tahiti, French Polynesia\\
  \inst{3} Ecole Centrale de Lyon, 69134 Ecully, France
}


\begin{document}

\maketitle

\begin{abstract}
Slow crack growth in a model of homogenous brittle elastic
material is described as a thermal activation process where stress
fluctuations allow to overcome a breaking threshold through a
series of irreversible steps. We study the case of a single crack
in a flat sheet for which analytical predictions can be made, and
compare them with results from the equivalent problem of a 2D
spring network. Good statistical agreement is obtained for the
crack growth profile and final rupture time. The specific scaling
of the energy barrier with stress intensity factor appears as a
consequence of irreversibility. In addition, the model brings out
a characteristic growth length whose physical meaning could be
tested experimentally.
\end{abstract}

\section{Introduction}
Although tensile rupture of atomic bonds requires a stress
comparable to the Young's modulus, brittle solids commonly rupture
at a much lower applied stress (typically 3 orders of magnitude
lower). Griffith's pioneering work \cite{Griffith} has clarified
the origin of this apparent weakening, postulating that small
cracks usually preexist in real solids, with stress concentration
at the crack tip strongly enhancing rupture. A somewhat similar
and quite striking effect is the occurrence of failure even though
the solid is stressed below its experimental breaking threshold
(i.e., even if stress concentration due to flaws is taken into
account). The physical process, sometimes referred to as
subcritical rupture, leads to a delay in the time for complete
macroscopic failure of the solid, with a strong dependence on the
amplitude of the applied stress.

A possible driving mechanism for subcritical damaging processes is
thermal activation as supported by the early experimental work of
Brenner and Zhurkov \cite{Brenner,Zhurkov}. Zhurkov introduced a
kinetic concept of strength of solids, where time to rupture
follows an Arrhenius law with an energy barrier decreasing with
increasing stress \cite{Zhurkov}. Interestingly, there is still
debate about whether temperature fluctuations might be sufficient
or not to nucleate microcracks and trigger crack growth. Recent
theoretical works \cite{Roux,Nattermann,Scorretti,Ciliberto} have
emphasized the effect of disorder in decreasing the effective
energy barrier (conversely increasing the "effective
temperature"). Other authors \cite{Blumberg,Sethna} have used
equilibrium statistical thermodynamics to study how cracks might
naturally appear from thermal fluctuations in an otherwise
homogenous material which goes into a metastable state when
stretched. However, another fundamental ingredient not usually
taken into account is the irreversibility of the rupture
mechanism, leading to dynamics out of equilibrium.

Temperature fluctuations in real materials are usually considered
too weak to be able to overcome a potential energy barrier
estimated as the free energy cost to reach Griffith's critical
crack length. But, the energy cost would be much smaller if
rupture was considered as an irreversible process, where thermal
fluctuations need only to be able to break atomic bonds one after
another. Irreversibility has already been taken into account by
Golubovic \etal \cite{Golubovic}, with the introduction of a
minimum crack opening beyond which the crack can not heal back.
However, Golubovic introduces a complex behaviour of stepwise
growth involving several bonds at a time, and eventually find that
lifetime is dominated by a critical crack length in the same
fashion as predicted directly from Griffith's criteria, albeit
with a smaller critical length and a different power law
dependence on applied stress. In this paper, we present a
different approach by choosing ab initio a distribution of stress
fluctuations directly linked to thermal fluctuations. Since our
goal is to describe irreversible crack growth, we are considering
the general case of a preexisting initial crack in a flat sheet of
material. In order to clarify our approach, we derive first the
energy barrier corresponding to our geometry in the framework of
Griffith's energy concept. Then, we give analytical solutions of
our model which yield the complete growth dynamics for the case of
a single crack. Finally, we compare the predictions of our model
to the results obtained with a numerical simulation of a
bidimensional elastic system in antiplane deformation.

\section{Energy barrier from Griffith's energy concept}
Griffith's prediction of a critical crack size beyond which there
is rupture, i.e. irreversible and fast crack growth, is derived
from a potential energy taking into account elastic energy due to
applied stress $\sigma$ and surface energy $\gamma$ needed to open
a crack as a function of a unique order parameter, the crack
length $\ell$. For a bidimensional geometry consisting of a flat
sheet with a crack perpendicular to the direction of stress, the
potential energy per unit thickness of the sheet reads:
\begin{equation}\label{Griffith}
 U =  - \frac{{\pi \ell ^2 \sigma ^2 }}{{4Y}} + 2\gamma \ell + U_0
\end{equation}
where $Y$ is the Young modulus and $U_0$ is the elastic energy in
the absence of crack. This energy reaches a maximum for a critical
crack length $\ell_c$ beyond which no stable state exist except
the separation of the solid in two broken pieces. From this
expression of the energy, it is clear that a stressed solid
without crack is in a metastable state with a lifetime depending
on the possibility to nucleate a crack with critical length.

Many authors \cite{Sethna,Golubovic,Pomeau1} have used models
essentially inspired by Griffith's energy concept to show that the
lifetime should follow an Arrhenius law with an energy barrier
scaling as $\Delta U\sim\sigma^{-2}$ in a bidimensional geometry
and $\Delta U\sim\sigma^{-4}$ in a three-dimensional one.
Following these authors, but taking into account the preexistence
of a stable initial crack of length $\ell_i$ and introducing
stress intensity factors $K_i = \sigma \sqrt{ \frac{\pi}{2}\ell
_i}$ and $K_c = \sigma \sqrt{ \frac{\pi}{2}\ell _c}$ , the energy
barrier becomes :

\begin{equation}\label{Griffith2}
  \Delta U = U(\ell_c)  - U(\ell_i)
  =\gamma \ell _i \left( {\frac{K_i}{{K_c }} - \frac{{K_c }}{K_i}}\right)^2
  = \gamma \ell _i f\left( K_i, K_c \right)
\end{equation}

Note that the energy barrier is a linear function of the initial
crack length multiplied by a function depending only on stress
intensity factors. This choice for the energy barrier implicitly
assumes that there is a possibility for the crack to explore
reversible states of crack length between initial and critical
crack length, and cannot be a good description if irreversibility
of crack growth is to be taken into account.

\section{Energy barrier for an irreversible crack growth}

Let us reflect on the fact that thermal fluctuations induce stress
fluctuations in the material. The uniaxial loading state of an
homogeneous solid at fixed temperature is described by its free
energy density: $\varphi(\sigma) = \sigma^2/2 Y$. Treating stress
as a fluctuating internal variable in a fixed volume $V$, the
probability to find a given stress is proportional to a Boltzmann
factor $\exp(-\varphi V/kT)$. Expanding free energy about an
equilibrium position $\sigma$, the distribution of stress
$\sigma_f$ is :
\begin{equation}\label{distribution}
  p(\sigma_f) \simeq \frac{1}{\sqrt {2\pi \langle \Delta \sigma\rangle^2}} \exp \left[ -\frac{\left(\sigma_f-\sigma\right)^2}{2\langle \Delta \sigma\rangle^2} \right]
\end{equation}
where $k$ is Boltzmann constant, $T$ temperature and $\langle
\Delta \sigma\rangle^2 = kT/(V\partial^2 \varphi/\partial
\sigma^2)=kT Y/V$\cite{Diu}. When a crack is present, stress
concentration increases the probability that breaking occurs at
the crack tip rather than anywhere else. We assume that stress
distribution at the crack tip remains the same as
eq.~(\ref{distribution}) despite the strong divergence of stress
and the breakdown of linear elasticity. Since the stress intensity
factor $K \approx \sigma \sqrt{\ell}$ gives a measure of stress
intensity close to the crack tip for a crack with length
measurement $\ell$ when external load is $\sigma$, we choose to
work directly with this quantity. The threshold for rupture at the
crack tip will be given by a critical value of stress intensity
factor $K_c$ as is usual laboratory practice. The cumulative
probability that stress intensity is larger than a given value
$\eta$ is then : $P(\eta ) = \int_\eta ^\infty {p\left(x\right)
\upd x}$. Since breaking is assumed to be an irreversible process,
the typical velocity of crack should be directly proportional to
the probability to have fluctuations giving stress intensity
larger than $\eta=K_c$ :

\begin{equation}\label{motion}
 \frac{\upd\ell }{\upd t} =  V_0 P(\eta=K_c) \simeq V_0 \sqrt {\frac{EkT}{2\pi}} \frac{1}{K_c  - K}\exp \left[ - \frac{\left(K_c  - K\right)^2}{2EkT} \right]
\end{equation}
where E is a dimensional constant proportional to the Young
modulus. In eq.~(\ref{motion}), the last equality is valid as long
as $ kT \ll \eta^2/2 E$. We also introduced a typical velocity
$V_0$ which represents the crack velocity when the condition for
crack advance is verified at all times $(P=1)$. This quantity is
typically the ratio of a microsopic scale (interatomic distance)
and a characteristic time (inverse vibrational frequency). Since
$K$ is a function of crack length $\ell$ , eq.~(\ref{motion}) is a
differential equation for crack evolution. To solve this equation
requires additional approximations since the dependence of stress
intensity factor on crack length is non-linear. First, we
introduce a reduced crack length $\phi \equiv
(\ell-\ell_i)/(\ell_c-\ell_i)$ to measure crack evolution as it
grows from its initial value $(\phi=0)$ to its ultimate stable
value $(\phi=1)$. Then, stress intensity factor can be written:
\begin{equation}
  K \approx \sigma \sqrt{\ell} = \sigma \sqrt{\ell_i+(\ell_c-\ell_i)\phi}
  \simeq K_i \left[1+\frac{1}{2}\left(\frac{\ell_c}{\ell_i}-1\right)\phi\right]
\end{equation}
where the last equality is a reasonable approximation giving less
than a $2\%$ error on stress intensity factor as long as $\phi
<1/2$ and $\ell_c < 2 \ell_i$. Another approximation will be to
take $K=K_i$ in the pre-factor of the exponential, because
neglecting the variation in stress intensity factor leads only to
a logarithmic correction of the crack velocity. As a consequence
of the last approximation, the crack velocity will tend to be
underestimated. Solution of the differential equation
(\ref{motion}) is then :
\begin{equation}\label{growth}
  t = \tau \left[1 - \exp \left( -\frac{\phi}{\phi_c}\right)\right]
\end{equation}
where $\tau$ gives the lifetime of the sample before fast rupture:
\begin{equation}\label{lifetime}
  \tau  = \tau_0\exp \left[ \frac{(K_c  - K_i )^2 }{2EkT} \right]
\end{equation}
with
\begin{equation}\label{tau0}
  \tau_0=\frac{2\sqrt{2 \pi E k T}}{K_i}\frac{\ell_i}{V_0}
\end{equation}
and $\phi_c$ is related to a characteristic growth length
$\lambda$:
\begin{equation}\label{phic}
  \phi_c  = \frac{\lambda}{\ell_c-\ell_i}=\frac{2EkT}{K_i(K_c-K_i)} \frac{\ell_i}{\ell_c-\ell_i}
\end{equation}

Note that the crack velocity : $\frac{\upd\ell}{\upd t} =
\lambda/(\tau-t)$, diverges as time comes closer to lifetime
$\tau$, which simply means that when time $\tau$ is reached slow
crack growth due to thermal activation is no longer the driving
mechanism, and a crossover towards fast dynamic crack propagation
will occur. The lifetime $\tau$ appearing in eq.~(\ref{lifetime})
follows an Arrhenius law with an energy barrier function of
initial and critical stress intensity factors similar to
eq.~(\ref{Griffith2}), but instead, there is no additional
proportionality to initial crack length. A similar scaling for the
energy barrier was found by Marder \cite{Marder}.

\section{Results from the simulation of a 2D elastic spring
network}

In order to check predictions from the previous analysis, we model
a bidimensional elastic system as a network of springs forming a
square lattice whose nodes can move only along an axis
perpendicular to the undeformed plane of springs. The elastic
restoring force of the spring is proportional to the variation in
displacement along the moving axis. This is a simplified model of
an antiplane deformation. A constant force is applied at two
opposite sides of the lattice, the direction of the force being
parallel to the moving axis, but reversed from one side to the
other. Starting from an equilibrium configuration, an external
force obtained from a normal distribution is applied in parallel
to each spring. A new static equilibrium configuration is
determined to find the fluctuations of spring forces. In this
procedure, fluctuations are quasi-statically coupled to a
temperature bath. Making the elastic constant between force and
displacement equal to unity, the temperature coefficient $kT$ is
numerically identical to the variance of spring force
fluctuations. Whenever the force on a spring exceeds a breaking
threshold $f_c$, the spring is cut and, since it is never
repaired, the process is irreversible. The crack itself is
modelled as a series of broken parallel springs, with the crack
direction parallel to the sides where the constant force is
applied. The size of the square lattice ($100\times100$ springs)
is determined to reduce finite size effects and obtain the correct
scaling of stress intensity factor with applied stress and crack
length. Time scale in the simulation is given as the constant time
between two configurations of force fluctuations in the system,
and length scale as the distance between two nodes of the lattice.

\begin{figure}
\onefigure[width=9cm]{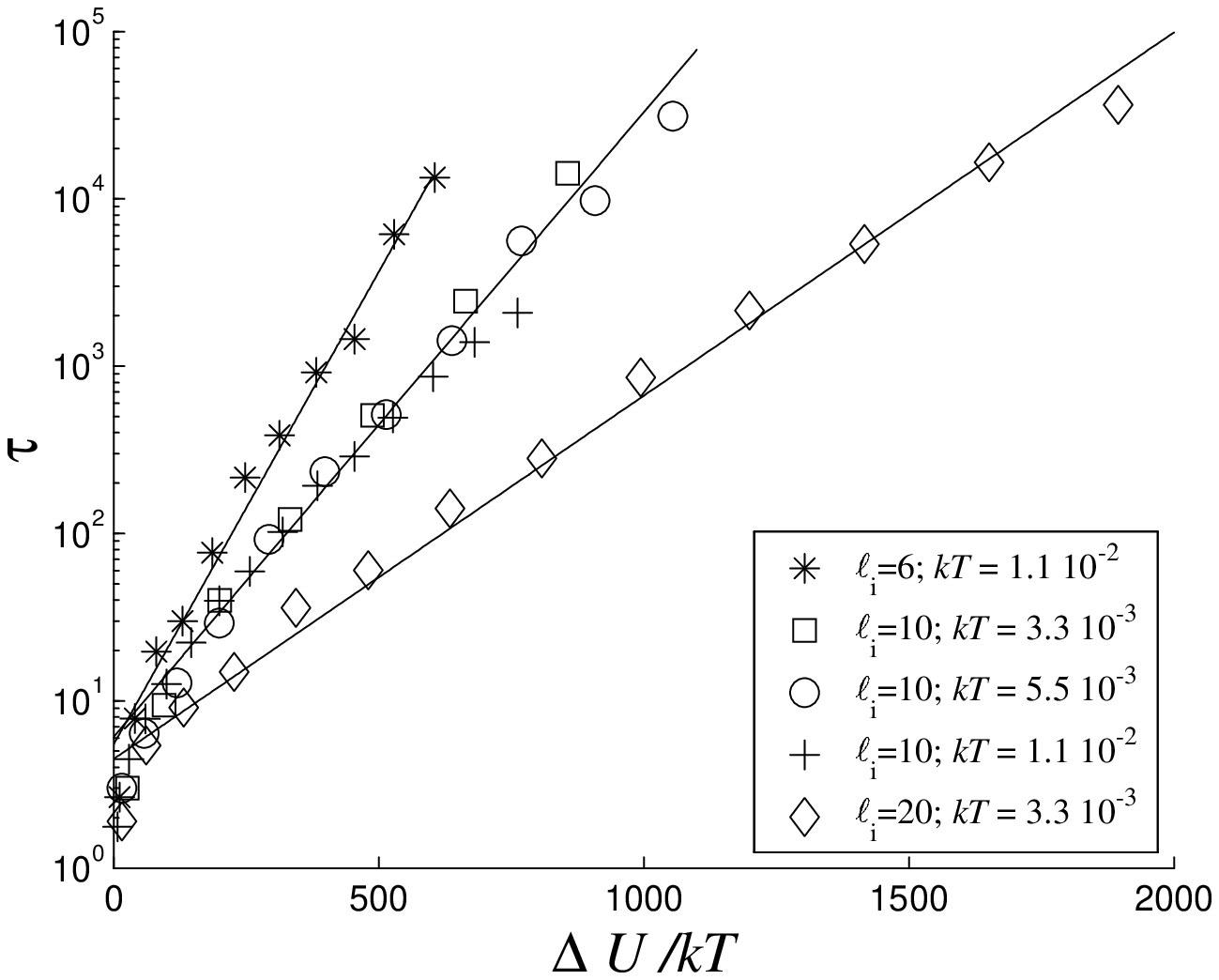} \caption{Logarithm of lifetime as
a function of the energy barrier as predicted by
eq.~(\ref{Griffith2}). Failure of data to scale with initial crack
length is the main observation. Straight lines are a guide for the
eye.} \label{fig.Lifetime1}
\end{figure}

\begin{figure}
\onefigure[width=9cm]{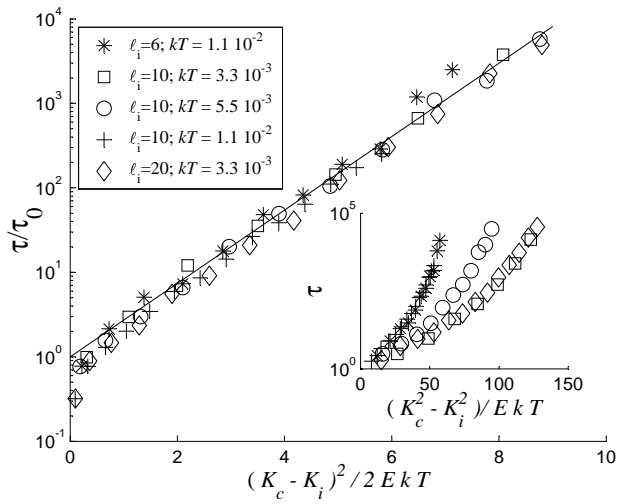} \caption{Logarithm of lifetime as
a function of the energy barrier as predicted by
eq.~(\ref{lifetime}). Scaling works for various temperature and
initial crack length. The straight line, slope $1$, is prediction
from eq.~(\ref{lifetime}). The inset shows failure of scaling with
energy barrier as predicted by eq.~(\ref{GriffithIrrev}).}
\label{fig.Lifetime2}
\end{figure}

The distribution of lifetimes $\tau$ obtained in the simulations
is a decreasing exponential with a typical width $\sqrt{\langle
\Delta \tau^2 \rangle}/\tau \approx 0.5$. In
fig.~\ref{fig.Lifetime1} and ~\ref{fig.Lifetime2}, we plot the
lifetimes averaged over an ensemble of 10 to 50 simulations for
samples with different initial crack lengths and different
temperatures. In fig.~\ref{fig.Lifetime1}, the logarithm of the
lifetime is plotted as a function of the energy barrier given by
eq.~(\ref{Griffith2}) and scaled by temperature coefficient $kT$.
For a given initial crack length, a good scaling is obtained at
various temperatures. However, when the initial crack length is
changed, the scaling with the energy barrier does not work at all.
In fig.~\ref{fig.Lifetime2}, the logarithm of the lifetime scaled
by $\tau_0$ is plotted as a function of the energy barrier given
by eq.~(\ref{lifetime}). This energy barrier scaling appears to be
correct whatever is the initial crack length or temperature. It
seems that the lack of dependence on initial crack length in the
energy barrier given by eq.~(\ref{lifetime}) is a signature of
irreversibility. Indeed, if one considers the potential energy
landscape obtained from eq.~(\ref{Griffith}), we can describe a
growth process where small irreversible steps $\delta$ occur, with
the energy barrier to overcome each of these steps proportional to
the derivative of the potential energy. It can be seen from
eq.~(\ref{Griffith}) that this energy barrier would be expressed
only as a function of stress intensity factor as in
eq.~(\ref{lifetime}). For each step, the energy barrier can be
written $\Delta U=\frac{\upd U}{\upd \ell}\delta=\left(K_c^2 -
K^2\right)\delta/Y$, and the waiting time for the crack to advance
with step $\delta$ is given by an Arrhenius law. Integrating the
time it takes to go from $\ell_i$ to $\ell_c$ gives in the limit
$\delta \ll \ell_i$ the following scaling relation:
\begin{equation}\label{GriffithIrrev}
  \tau\approx\exp\left(\frac{K_c^2 -K_i^2}{Y k
  T}\delta\right)
\end{equation}
Although this approach does correct the problem of scaling with
$\ell_i$, it completely fails to give the correct scaling of the
energy barrier with stress intensity factors; furthermore, for a
given initial crack length, temperature scaling is lost as can be
seen from the insert in fig.~\ref{fig.Lifetime2}.

\begin{figure}
\onefigure[width=9cm]{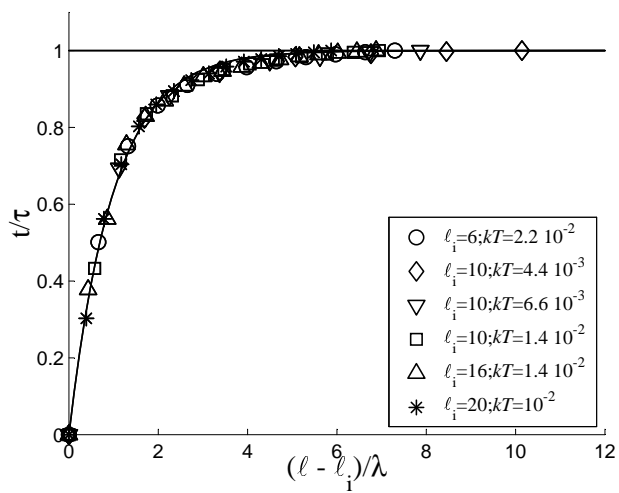} \caption{Rescaled time as a
function of rescaled crack length for various initial crack
lengths and temperatures. For each set of data, $\lambda$ is
extracted as the only fit parameter since lifetime is known. Solid
line corresponds to eq.~(\ref{growth}).} \label{fig.Growth}
\end{figure}

\begin{figure}
\onefigure[width=9cm]{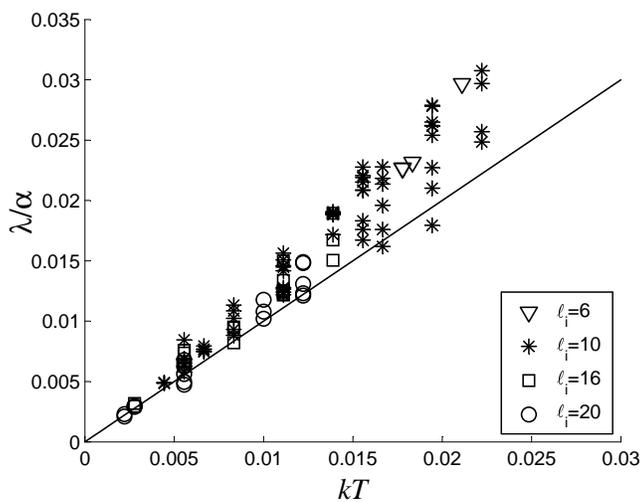} \caption{Linear dependence of
$\lambda/\alpha$ with temperature coefficient $kT$ for various
initial crack lengths and applied forces. The solid line shows the
behaviour expected from the model (slope $=1$).}
\label{fig.Lambda}
\end{figure}

To determine the characteristic growth length $\lambda$ which
appears in our model, it is necessary to look at the complete
growth dynamic. As for lifetimes, there is a strong dispersion in
growth curves obtained for different simulations with the same
temperature and initial crack length. In order to get statistical
information about crack growth, we determine the average time it
takes for the crack to reach a certain length. In
fig.~\ref{fig.Growth}, which shows flowing time as a function of
crack length, the exponential behaviour obtained in
eq.~(\ref{growth}) is recovered and yields a value of $\lambda$.
In fig.~\ref{fig.Lambda}, we plot $\lambda$ scaled by the
coefficient $\alpha=2E\ell_i/[K_i(K_c-K_i)]$ as a function of
$kT$. The linear dependence of $\lambda$ on temperature predicted
by eq.~(\ref{phic}) appears clearly for various initial crack
lengths and applied forces. In addition, the correct rescaling of
data with initial crack lengths is mainly related to the linear
dependence of $\lambda$ on $\ell_i$. Dispersion of data for a
given temperature and initial crack length occurs systematically
above prediction of the model (see solid curve in
fig.~\ref{fig.Lambda}). This can be understood as a consequence of
the approximations in our model underestimating crack velocity.
This interpretation is consistent with the increase in dispersion
of the data as $\ell_c$ increases.

\section{Conclusion}

We have shown that by taking into account distribution of stress
fluctuations, we can describe thermally activated and irreversible
crack growth and obtain the correct scaling of lifetime with
applied stress and initial crack length. Previous predictions
based on Griffith's potential energy
\cite{Sethna,Golubovic,Pomeau1} fail to take into account
irreversibility of the rupture process, thus do not apply to our
configuration. Our challenge is to find a way to include
irreversibility in a problem of total potential energy
minimization, since simple arguments fail to give the correct
behaviour. More work should be done to check that stress
divergence at the crack tip does not preclude the derivation of
stress distribution presented in this paper. Direct comparisons
with experiments can be performed to test the relevance of the
characteristic length scale for crack growth introduced in our
model, as well as scaling of both lifetimes and characteristic
growth length on initial crack length.

\acknowledgments

\newpage

\newpage

\end{document}